# A Systematic Mapping Study on Dynamic Metrics and Software Quality

Amjed Tahir and Stephen G. MacDonell

*SERL, School of Computing & Mathematical Sciences*
*Auckland University of Technology*
*Auckland, New Zealand*
amjed.tahir@aut.ac.nz, stephen.macdonell@aut.ac.nz

**Abstract**

*Several important aspects of software product quality can be evaluated using dynamic metrics that effectively capture and reflect the software's true runtime behavior. While the extent of research in this field is still relatively limited, particularly when compared to research on static metrics, the field is growing, given the inherent advantages of dynamic metrics. The aim of this work is to systematically investigate the body of research on dynamic software metrics to identify issues associated with their selection, design and implementation. Mapping studies are being increasingly used in software engineering to characterize an emerging body of research and to identify gaps in the field under investigation. In this study we identified and evaluated 60 works based on a set of defined selection criteria. These studies were further classified and analyzed to identify their relativity to future dynamic metrics research. The classification was based on three different facets: research focus, research type and contribution type. We found a strong body of research related to dynamic coupling and cohesion metrics, with most works also addressing the abstract notion of software complexity. Specific opportunities for future work relate to a much broader range of quality dimensions.*

**Keywords** - dynamic metrics; software metrics; dynamic analysis; mapping study; software quality

## I. INTRODUCTION

Measurement plays a crucial role in contemporary software development, deployment and use by enabling managers and software engineers to efficiently evaluate their software products. In doing so two different sets of software metrics can be used: static and dynamic. Static metrics are the class of measures that capture the static properties of software components. As the name implies, this group of metrics does not need the software to be executed in order to determine their values. Although these measures are certainly useful, they cannot reflect key characteristics (such as dynamic dependencies between software components) of the running software system. Such characteristics are therefore measured by a growing body of dynamic metrics. Typically this occurs during software execution (i.e., at runtime) although, in some cases, simulation can be used instead of actual execution.

The aim of this work is to systematically investigate the emerging body of research on dynamic software metrics to identify issues associated with their selection, design and implementation. To the best of our knowledge, only one review [1] has been conducted to date in the field of dynamic metrics. However, there are other relevant reviews that have been conducted in similar areas, such as software metrics in general [2-3], and program comprehension [4]. This paper therefore presents the results of an intensive and recent mapping study to identify and classify current research in the field of dynamic software metrics. The objectives of this review are to provide a general overview of the research activities in the field and to guide researchers and readers regarding potential research gaps that can be further studied.

The remainder of this paper is organized as follows. *Section II* provides a background to the field and presents some relevant surveys. *Section III* explains our review methodology with a description of our review protocol, questions and process. Results of the study are presented in *Section IV* followed by a discussion of the key findings in *Section V*. *Section VI* outlines the relevant threats to validity; finally, *Section VII* concludes the paper.

## II. BACKGROUND

### A. Dynamic Metrics

The fields of dynamic metrics and dynamic analysis are naturally interrelated. Dynamic analysis – "the analysis of data gathered from a running program" [4, p.684] – reveals traits of software components during execution, and dynamic metrics are used to measure specific runtime properties of programs, components, subsystems and systems. Thus, we may consider the collection of dynamic metrics as a part of the dynamic analysis paradigm. The concept of dynamic analysis has been explained in detail in prior literature [4-6].

The collection of dynamic metrics can be accomplished in different ways. Most common is to collect the data by obtaining execution trace using dynamic analysis techniques during software execution. Another method is to simulate runtime behavior based on executable modules and interaction diagrams (such as in UML and Real-time Object Oriented Modelling (ROOM) languages). Each of these techniques has advantages and disadvantages. The first approach provides *actual* figures reflecting system behavior, as it captures the true values at runtime. A disadvantage of this approach is that it is only feasible in the later stages of development. On the other hand, simulation does not require executable code and the metrics data can be collected at an earlier stage. However, given likely changes between design and code, this technique is not as accurate and precise as its execution-based counterpart. Despite these difficulties, both techniques have been examined empirically to collect and test several dynamic metrics [7-11].

As stated, the focus of this paper is on dynamic software metrics techniques and applications and so, beyond a brief comment here, we do not concern ourselves with static metrics. Static analysis (and measurement) depends on artifacts such as source code and documentation [5, 12-13]. In principle it can support complete code coverage by considering all possible scenarios and paths in the software. It is also generally acknowledged that static metrics are easier to collect and/or compute compared to their dynamic counterparts. In addition, static metrics can be verified by inspection of the program source code. Nevertheless, this process of source code inspection may not be sufficient to capture all the properties of interest, and verification can be a difficult process [9]. Empirical studies show that static measurement is – unsurprisingly – not sufficient for capturing dynamic dependencies among system modules such as those related to polymorphism, dynamic binding and inheritance [7]. It is also noted that static analysis may result in a huge amount of data [6] that can be difficult to understand and summarize [14].

Data gathered from dynamic analysis provide broader and more precise coverage than static analysis in capturing runtime behaviors [6, 15-16]. Dynamic analysis requires instrumenting a program to examine aspects of interest. This instrumentation can help a developer to collect precisely all the information needed to address a particular problem in the software [5]. Dynamic analysis has been used largely in the fields of software measurement [7, 14, 17-18], software maintenance and reengineering [19-20], in clustering [21], and in program understanding and comprehension [12, 22].

Due to the fact that dynamic analysis supports the collection of the information needed to address a particular problem based on its actual appearance during execution, developers are supported in relating changes in program inputs to changes in internal program behavior and resulting outputs [5]. Another advantage of the use of dynamic metrics is the level and focus of detail available. Examining a specific execution scenario dynamically limits the scope of investigation, which provides more detailed and targeted results about the specific execution scenario [6].

The reported disadvantages of the technique lie in two different but related points: incompleteness and limited generalization. Incomplete coverage is one of the major arguments against the use of dynamic analysis, as the data gathered may be broad but shallow, reflecting only the scenario that was executed [13]. However, it has been argued equally strongly that incomplete coverage of software code is not necessarily a weakness [15]. To understand a program's behavior we do not always need complete information about it; rather, evaluators need sufficient information that can help them to form concepts about the software's structure. The incomplete coverage 'problem' leads to another cited disadvantage: limited generalization of the results obtained. Dynamic metrics' results may not generalize for future executions, given that the gathered data pertain solely to the scenario that was executed at a given point in time [13-14]. There is no way of assuring that the scenario(s) in which the program was run is (are) representative of all possible program scenarios and executions [6]. One of the obvious ways to minimize the negative impacts of this problem is to execute a range of scenarios that represent major execution paths (rather than all possible execution paths) according to cost-benefit. The combined results compiled across multiple scenarios and paths can help to provide more complete as well as more accurate results. This could help software engineers to sufficiently generalize their findings and results to other important components of the software system.

## B. Related Reviews

According to our review of the relevant literature the first works on dynamic metrics were those of Voas [23] and Munson and Khoshgoftaar [24]. Both papers were published in 1992, and focus on dynamic complexity metrics. The body of literature has grown steadily since that time, leading to a small number of survey studies in recent years.

Most closely related to our work is the study of Chhabra and Gupta [1] who conducted a review to address research problems, challenges and opportunities in the dynamic software metrics domain. Their paper discussed several notable works in the field. This was an informal review as there were no defined research questions, search process or data extraction process. As a result the review did not cover some important works in the area such as [9, 14, 17, 23, 25-27], and the latest article considered in the review had been published in 2006. We have found a number of studies (31 articles) on the topic that were published between the latest article cited in Chhabra and Gupta's study and our study.

In 2009 Cornelissen *et al*. [4] published a systematic review concerning program comprehension as achieved through dynamic analysis. Their paper considered the limitations of prior work and opportunities for further improvement that could be contributed to the area. However, their focus was mainly on the dynamic analysis *techniques* rather than metrics and measurement. Another more general review to systematically inspect software metrics from project, process and product points of view was reported by Gomez *et al.* in 2008 [2]. Their work intended to answer the high-level questions of what, when and how software should be measured. Kitchenham [3] conducted another general mapping study to identify influential software metrics papers published between 2000 and 2005. She also investigated the possibility of using secondary studies to integrate research results. While product metrics were considered by these two reviews no specific attention was given to dynamic metrics.

## III. METHOD

Systematic mappings have been recommended for research areas where there is an emerging body of relevant primary studies. Mapping reviews aim to identify and characterize all research works that are related to a specific topic, using a defined and defendable search strategy. In this work we are targeting the area of dynamic software metrics, both their design and current use. (Note that in this work we are using the systematic review guidelines suggested in [28].) The result of a mapping study is a set of papers related to a topic area categorized according to a variety of dimensions, and counts of the numbers of papers in those various categories [29]. These results can help to highlight issues that warrant further investigation in primary studies.

The reminder of this section describes our review protocol, questions and process.

### A. Review Protocol Overview

Our review protocol is shown in Fig. 1 and represents a general overview of our review study. Only the major steps of our study are shown. Details of the protocol elements are addressed in the following subsections.

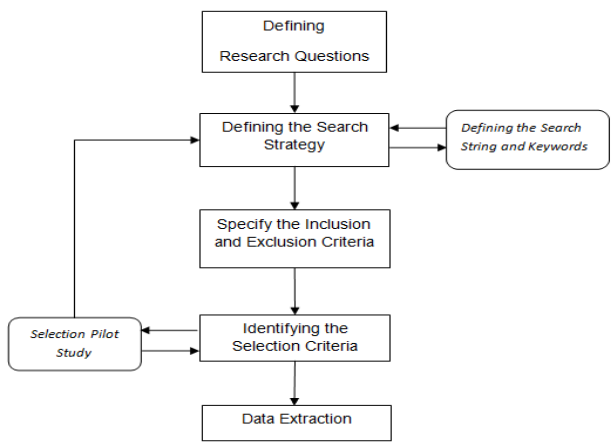

Figure 1. Review protocol overview.

### B. Review Research Questions

For any review, defining one or more appropriate research questions is a critical step. Based on the review questions researchers can determine appropriate search and data extraction strategies. In this work we intend to answer the following two key questions:

*RQ1* Which aspects of dynamic metrics have been most frequently subjected to study?

*RQ2* Which aspects of dynamic metrics could be recommended as topics for future research?

*RQ1* addresses recent and current research into dynamic metrics in software development, categorizing all research activities in the field. It is essential to know what metrics have already been developed and the characteristics these metrics actually measure. Answering this question will contribute to determining the usefulness and the drawbacks of this group of metrics. By knowing the metrics, their coverage and their mechanisms of action, we should be able to identify any current difficulties as a precursor to suggesting solutions or possible avenues of further primary investigation – thus informing *RQ2*. *RQ2* is expected to be of help in directing future research in the field based on stated and implied research problems and gaps.

### C. Search Strategy

We divided our search into two main phases: *Automatic* and *Manual*. An automatic approach is used to search for materials via electronic search engines using a pre-defined and tested search string. A manual search, on the other hand, is performed by one or more reviewers reading through targeted publications (normally journals and conference proceedings). Such a combined approach should cover a wider range of materials than would be possible with one approach, reducing the likelihood of the review missing relevant literature. Our search was conducted for the period between January 1992 and June 2011, as the first two papers to study dynamic metrics were both published in 1992.

#### 1) Automatic Search

We conducted our automatic search using two different electronic sources, namely: *SCOPUS* and *Google Scholar*. In order to have efficient and close-to-complete coverage of peer-reviewed and published works we first used the *SCOPUS* search engine. *SCOPUS* provides web-based access to over 4,000 international publishers including 531 open access journals. *SCOPUS* also indexes databases that publish much of the relevant software engineering literature, including: *IEEE*, *ACM*, *Elsevier*, *Springer* and *Wiley-Blackwell* publishers. The *Google Scholar* search engine was used to look particularly for grey literature and articles that could not be found by the *SCOPUS* search engine, or were not published by the abovementioned publishers. It is important to highlight that *Google Scholar* was used here mainly as a secondary source to improve the level of assurance regarding coverage of the relevant literature.

After we selected the relevant electronic sources we defined the search string (*Fig. 2*) to be used in our automatic search process. It has been noted in prior reviews that using a specific and verified search string is likely to improve the search process and increase the likelihood of finding relevant studies and also reduce search workload [30]. We applied our search string to search for materials via the *SCOPUS* database portal. Unfortunately, the nature of *Google Scholar's* search structure did not support us effectively using our search string as defined. When we attempted to do so it returned a huge, unworkable number of papers and materials, many of which were not even related to the field. We therefore moved to use a very simple search string term, "*Dynamic Metrics*", to search for papers via *Google Scholar*.

> *((software OR program) AND ("dynamic metrics" OR "dynamic metric" OR "dynamic measurement" OR "runtime metrics" OR "dynamic measure")) OR ("dynamic analysis" AND ("program comprehension" OR "program understanding") AND metrics)*

**Figure 2.** Search string defined for the automatic search.

*2) Manual Search*

Unless an automatic search string is extremely obscure it is basically a given that such a search will find more results in comparison to a targeted but comparatively labor-intensive manual search. However, if they are not conducted with care, automatic searches can be of poor quality [31].

The value of our manual search is in increasing the reliability of our search process by ensuring that we do not miss important literature in the field that cannot be found using the search string. This happens mainly due to the restriction criteria of automatic searches. Combining these two techniques can thus solve problems that we might face if we conducted either the manual or automatic search only.

During this phase we searched manually for articles in a list of eight journals and nine conference and workshop proceedings relevant to our research topic. Based on our prior knowledge of the research domain, these journals, conferences and workshops were known to be closely related to software metrics, program analysis, and software maintenance fields. The full list of the selected journals and proceedings (for the manual search only) is shown in *Table I*.

TABLE I. MANUAL SEARCH VENUES AND RESULTS

| Journal | No. of Articles | Conference | No. of Articles |
|---|---|---|---|
| IEEE TSE | 3 | ICSE | 4 |
| ACM TOSEM | 0 | ICPC (IWCP) | 11 |
| JSS | 4 | METRICS | 2 |
| IEEE Software | 2 | ICSM | 3 |
| I&ST | 0 | PCODA | 5 |
| JSME | 1 | WETSoM | 1 |
| ESEJ | 0 | ISESE | 0 |
| SQJ | 1 | ESEM | 0 |
|  |  | WODA | 0 |
| **Totals** | **11** |  | **26** |

*3) Reference Checking*

In addition, the first author checked the lists of references in all the articles initially identified as relevant. This additional step can help reviewers to uncover a wider range of articles and further minimize the chance of omitting significant work in the field [4]. We conducted such a reference check on the final set of papers that remained in consideration after all the inclusion and exclusion tests.

### D. Study Selection Criteria

Based on the identified research questions and goals, we defined the following selection criteria to be applied to the candidate articles:

- All works must be strongly relevant to dynamic software metrics. Our goal is to select works focused on the design of dynamic software metrics and their application. This would exclude such topics as tracing, debugging and program slicing.

- As the review is focused on dynamic metrics topics (process, techniques, methods and tools), studies concerned with static metrics would not be considered.

### E. Inclusion and Exclusion Criteria

Inclusion and exclusion criteria are used to filter and rule out studies that are either not relevant to the defined review questions or not of the standard required to warrant attention. The review included papers published between January 1992 and June 2011. We included studies from the following domains: *Dynamic metrics and measures*, *Dynamic program analysis*, *Program comprehension and understanding*, and *Software re-engineering, evolution and reverse engineering*. We also included papers that focused on using dynamic metrics to assess and measure different software aspects.

We excluded the following:

- Studies not in English.

- Editorials, prefaces, covers, books, interviews, news, correspondence, comments, tutorials, readers' letters and summaries of workshops and symposia.

- Duplicate studies (e.g., several reports of the same study published in different places or on different dates). In such a case, the more detailed and comprehensive paper is selected (in most cases, journal versions were selected over conference versions).

- Papers that study dynamic analysis and performance profiling techniques without any consideration of dynamic metrics.

- Work mainly focused on static analysis and metrics.

### F. Search and Selection Pilot Study

Some authors recommend that a pilot study be conducted to validate the study selection approach and to refine/confirm the search strategy before conducting the actual full-scale review. The main purpose of the pilot study is to validate the search string, before applying it in a larger scale search. Our pilot study was conducted using a short, intentionally inclusive string (*software* AND *dynamic* AND *metrics*) to search for materials in the *IEEE Xplore* database. We decided to choose a shorter search period (between January 2001 and June 2011) to have more control over article selection. The total number of articles retrieved was 298. After applying our defined inclusion and exclusion criteria, we selected 22 relevant primary studies.

We validated the results of this pilot study based on article relevance as well as our familiarity with the field. In respect to the latter, we were able to identify 15 of the 22 articles found by the search as being familiar to us, lending a degree of confidence that our search would indeed (at least) find papers relevant to our research questions.

### G. Mapping Review Process

Our review process was composed of six main stages. *Fig. 3* highlights the review process and the number of publications under consideration at the end of each stage. As shown in *Fig. 3*, we conducted our automatic and manual searches in stage 1 of the process. We initially started with the automatic search in the *SCOPUS* database. Then we performed our manual search on the selected list of journals and proceedings. In stage 2, we applied the first filter by discarding papers with irrelevant titles that had been returned by the automatic search. In stage 3, we filtered articles based on the abstracts, but when the abstract was ambiguous, we also browsed the introduction of the paper. Stages 2 and 3 were not needed for the manual search, as the selection in the first place was based on the title and abstract. We then returned back to stage 1 and conducted the second automatic search, using *Google Scholar*. In stage 4 of the search process, we combined the results of both manual and automatic searches. Then, in stage 5, we performed a full text review of all the papers obtained. Results from stage 5 were added to a final list of papers in stage 6 of the process. After that we conducted a reference check on all selected papers in the final list. Additional papers were added to the final list and another round of reference checking was conducted on the newly added papers. Inclusion and exclusion criteria were applied at all stages. The final list included 60 papers. All the extracted data were stored in separate *MS Excel* data sheets for ease of analysis.

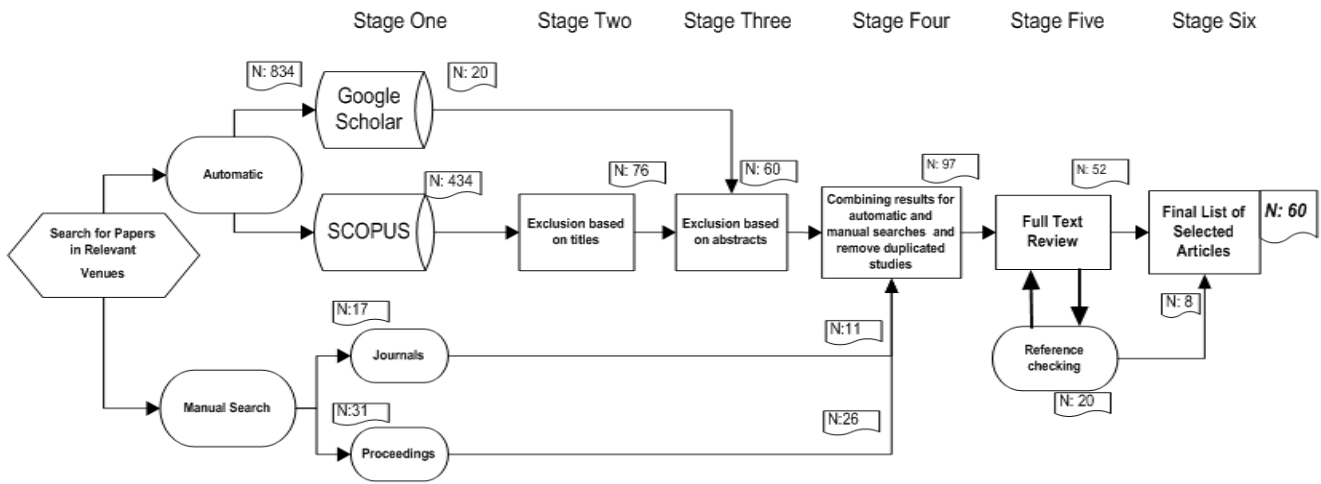

**Figure 3.** Mapping review process.

### H. Primary Study Classification Schemes

As the basis of classification in this work we used the schema presented in [32]. Publications were categorized according to the following schemes: research type, research focus and contribution type.

For the research type, we used the paper categorization proposed by Wieringa *et al*. [33]. This classification has been recommended in prior work [32] and has been used in recent systematic review and mapping studies, such as [29]. Research type is therefore categorized into the following [33]: ***Evaluation, Proposal, Validation, Philosophical, Opinion,*** and ***Personal Experience*** papers.

Research focus is classified into four main categories: (1) ***Estimation***: metrics that are used for the purpose of estimation (of errors, for instance); (2) ***Design level***: metrics that can be collected at the design stage or early in the development process; (3) ***Code level***: metrics derived from the source code; (4) ***Reengineering/comprehension***: metrics that are used for the purposes of reengineering, comprehension, understanding or maintenance.

Finally, contribution type comprises four main categories: (1) ***Method***: description on how to measure specific software aspects; (2) ***Process***: research that deals with the measurement process; (3) ***Tool***: any automated tool designed to support the measurement process; (4) ***Metrics***: metrics designed to measure runtime aspects of software.

## IV. MAPPING RESULTS

After determining all of the relevant articles the total number of selected primary studies was 60. (The numbers of articles found using manual searches were shown in *Table I*.) The distribution of the selected studies (*Table II*) shows that the majority were published in conference proceedings. As shown in *Fig. 4*, it is evident that the number of publications addressing dynamic metrics has generally increased from 2002 and that 32 of the 60 articles were published between 2007 and 2011. Of these, 2010 had the highest number of articles published in a single year.

As it is difficult to categorize review papers we decided to discard them from any further analysis (only one review paper was found). Therefore, 59 articles were further classified using the classification schemes described above.

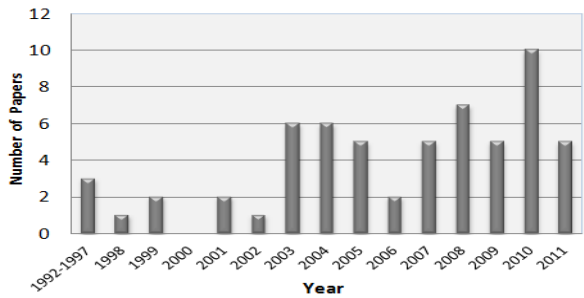

**Figure 4.** Articles distribution per year.

**TABLE II.** DISTRIBUTION OF ARTICLES PER SOURCE TYPE

| Publication Type | No. of Studies | Percentage |
|---|---|---|
| Journals | 18 | 30.0 |
| Conferences | 31 | 51.7 |
| Workshops | 7 | 11.7 |
| Technical Reports/ Newsletters | 4 | 6.7 |

In order to provide a more accessible representation of our results we summarized the data using tables and visual representations. Visual techniques, such as bubble charts, have been recommended previously for mapping data presentation [32]. *Fig. 5* depicts a map of publications over our defined classification criteria. Research focus is shown on the *Y* axis, Contribution type is shown on the right *X* axis, and Research type is shown on the left *X* axis. The size of each bubble represents the number of publications in the corresponding category pair. As is evident, the proposal and evaluation of code metrics and methods currently dominate the body of literature on this topic. Breaking *Fig. 5* down, *Fig. 6* and *7* show the distributions of articles per research and contribution type, respectively. Articles distributed by research focus are shown in *Fig. 8*.

Based on the retrieved data, we also found that a high proportion of the papers dealt with OO metrics: more than 75% of the studies. The remainder addressed a mix of procedural, aspect-oriented and service-oriented approaches.

Studies of specific, named metrics were also categorized based on their use. (Note: some works were not included in this classification as the particular metrics used were not specified.) It is important to note that many of the metrics and factors are interrelated. For example, coupling metrics are also often considered to be related to complexity.

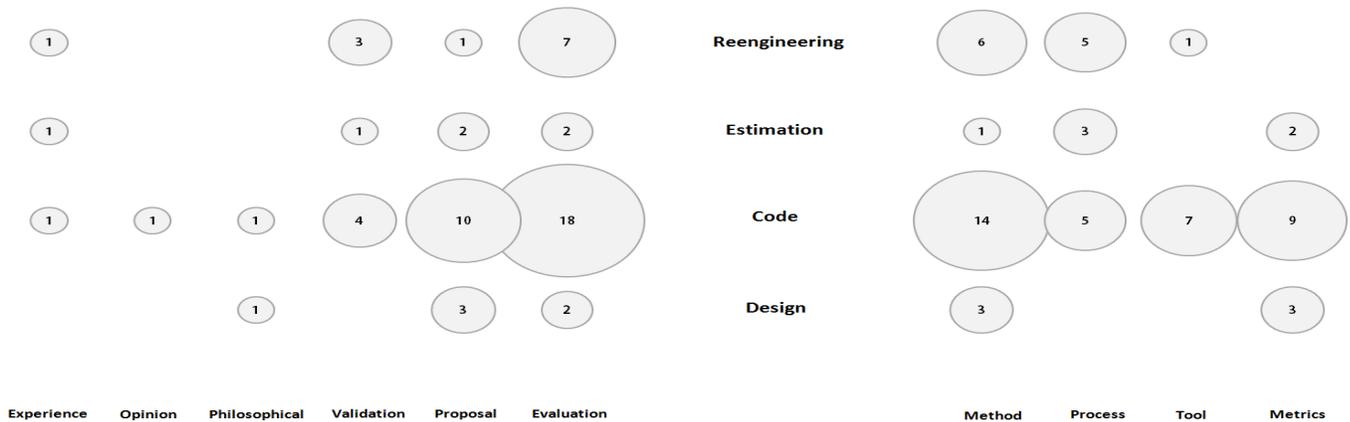

**Figure 5.** Bubble Chart Map of Research focus over Research and Contribution types.

However, we have distinguished between metric categories based on the stated (or in a few cases, presumed) intent of each study. The distribution of papers per topic is shown in *Table III*.

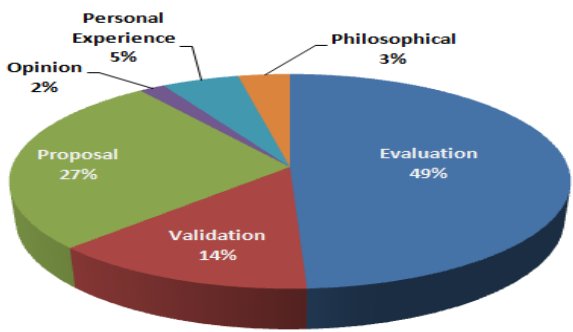

**Figure 6.** Articles distribution by research type.

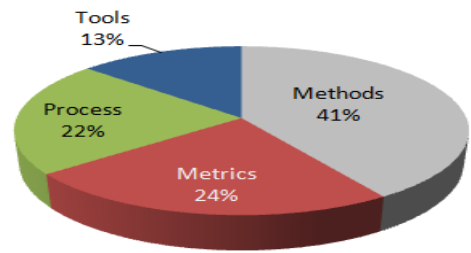

**Figure 7.** Articles distribution by contribution type.

## A. Coupling

We found a relatively strong body of research related to dynamic coupling [7, 11, 35, 41].

Of note is that most of the dynamic coupling measurement works were motivated by the Chidamber & Kemerer (C&K)

metrics suite [68] and their well-known static coupling measure *Coupling Between Objects* (CBO). Cho *et al.* [21] introduced a metric that assesses dynamic coupling at an object level by measuring the message passing load. Yacoub *et al.* [11] proposed two dynamic coupling metrics (i.e., *Import* and *Export* object coupling) to measure coupling at the design level using Real-time Object Oriented Modelling (ROOM) charts. These authors later applied the same set of metrics to estimate and assess reliability risks during early phases of development [34]. Arisholm *et al.* [7] introduced a set of code-level dynamic coupling metrics based on the dynamic analysis of systems. The authors found that dynamic coupling measures can be a good indicator of the complexity and change-proneness of a system. Burrows *et al.* [48-49] empirically examined several dynamic coupling metrics, in the context of AOP design. It was found that most of the existing AOP coupling metrics did not correlate well with several faults related specifically to aspect-orientation. Out of these metrics, the authors found that *Base-Aspect Coupling* (BAC) and *Crosscutting Degree of an Aspect* (CDA) were the two metrics that displayed the strongest correlation with faults [49]. Furthermore, the authors indicated that extensions of the C&K object-oriented metrics had not proven to be good indicators of fault-proneness in AOP.

TABLE III. METRIC COVERAGE IN DYNAMIC METRICS RESEARCH

| Metric type | Studies |
|---|---|
| Coupling | [7] [9] [10] [11] [14] [19] [21] [22] [27] [34] [35] [36] [37] [38] [39] [40] [41] [42] [43] [44] [45] [46] [47] [48] [49] |
| Cohesion | [14] [18] [21] [27] [38] [45] [50] [51] [52] |
| Complexity | [11] [23] [24] [34] [45] [53] [54] [55] [56] |
| Method invocation | [15] [17] [57] [58] [59] [60] [61] |
| Polymorphism | [9] [25] [62] [63] [64] |
| Memory-related | [9] [17] [27] [54] [65] |
| Code coverage | [9] [27] [35] [66] |
| Size/Structure | [9] [48] [65] [66] [67] |

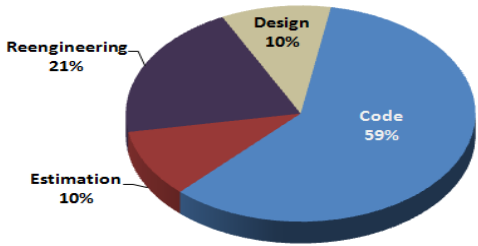

Figure 8. Articles distribution by research focus.

## B. Cohesion

Cohesion is another reasonably well-studied topic and, as with coupling, most of the proposed dynamic cohesion metrics are based on the C&K metrics suite [68].

One of the earlier works on runtime cohesion [50] uses a novel program execution-based approach to measure the module (functional) cohesion of legacy systems, by applying a dynamic slicing approach to overcome the limitations of over-estimation resulting from the classic static slice. Recent work of Gupta and Chhabra [18] defined a set of dynamic metrics to measure cohesion at an object level. They defined four types of metrics to measure four levels of relationships. Their empirical evaluation showed that these new measures were more accurate when compared to other existing metrics.

A runtime form of C&K's *Lack of Cohesion in Methods* (LCOM) metric was introduced in [51]. Two new metric variants are the *Run-time Simple LCOM*, which is derived directly from the C&K static LCOM metric, and the *Run-time Call-Weighted LCOM* metric, which measures each instance variable by the number of times it is accessed. Three dynamic measures were proposed in [52] based on the Read/Write interactions between methods. These metrics were also inherited from the C&K cohesion metrics. Cho *et al.* [21] measured cohesion based on the message passing load, taking into account both the number of messages as well as the load carried in each.

## C. Complexity

Like software quality, complexity is an amorphous concept that, when measured, must be operationalized in other terms. That said, the studies considered here utilized the term 'complexity' to refer to various aspects of code or design so we have retained it for this discussion (without attempting any reinterpretation of our own).

Munson and Khoshgoftaar defined their *Functional Complexity* metric, said to measure the dynamic complexity of systems [24]. This metric was further used in a later, related study [53] to estimate and examine the test effectiveness of software programs. Two additional dynamic complexity measures were introduced in their work, namely the *Fractional* and *Operational Complexity* metrics. The study [53] found a direct relationship between these dynamic metrics and software faults. Yacoub *et al.* [11] later used the *Operational Complexity* metric to measure the 'dynamic complexity' of software. This metric is based on the static *Cyclomatic Complexity* metric and can be collected during the early stages of development using *State Charts* designed using the ROOM modeling notation.

Voas [23] introduced the *Revealing Ability* dynamic metric, which purports to measure semantic software complexity by predicting a program's ability to allow faults to be undetected during dynamic random testing. Another run-time complexity metric was introduced by Mathur *et al.* [69] based on decision points in code, where one option is chosen from an available selection.

## D. Polymorphism

Polymorphism is a software characteristic that needs to be measured dynamically. Dufour *et al.* [9] used 17 different dynamic metrics to measure polymorphism from different perspectives. Choi and Tempero [63] presented their *Polymorphic Behaviour* metric. This metric was examined mainly in the context of software reusability. Sandhu and Singh [64] briefly described and then used eleven different metrics to measure polymorphism at various levels.

### E. Other Metrics

Space precludes discussion of all of the other metric type categories (*Table III*) but some specific examples are considered here. Burrows *et al*. [48] used metrics to measure code churn, which has been shown to have a direct effect on the incidence of faults. Cai [70] proposed a set of dynamic metrics that was used to measure the modularization of software components during maintenance tasks. This was achieved by comparing different versions of a program. The metrics considered the differences between versions, in terms of modules added, deleted, and changed.

A requirement-based dynamic metric was proposed in [8] that could be used to predict network communication loads. The authors contended that this metric could be applied during the early stages of development, using data collected from a system's requirement specification and defining a 'typical' usage scenario. Mendes *et al*. [67] empirically tested a size-based dynamic metric to measure the features and functionalities of web-based applications. This metric was used alongside other static metrics for the purpose of cost estimation during the early stages of development. Gani *et al*. [61] proposed a solution for dynamic metric collection, to support adaptation via object mobility, for mobile applications. Six different metrics were used to measure aspects related to execution frequency and time, performance, and resource utilization.

### F. Metrics Suites

Several works have introduced a set of dynamic metrics that could be used to collectively measure various aspects of software programs. These metrics could be collected either separately or as a set. In addition to broad coverage a key benefit of using a metrics suite is that there is potential for multiple measures of the same underlying construct [68].

Dufour *et al*. [9] proposed a set of dynamic metrics for *Java* programs that could be used to measure several runtime properties of software programs. These dynamic metrics were gathered into five main groups: size and structure of programs, data structures, polymorphism, memory, and concurrency. These metrics were examined empirically against several well-known *Java* benchmarks. The authors contend that these metrics could be used to capture relevant qualities; especially for compiler optimization developers.

A metrics suite for *Component Based Development* (*CBD*) was presented by Narasimhan and Hendradjaya [71]. Several metrics, both static and dynamic, were designed to measure the complexity and criticality of component assembly. Rothlisberger [60] designed and implemented five different dynamic metrics to enhance the *Eclipse IDE* analysis of Java applications, in order to help developers achieve a better understanding of their software. These metrics collect method execution and memory related data.

Several tools have also been implemented to help automate the dynamic measurement process. Some of these tools are implementations of the above metrics suites. A summary of these tools is shown in *Table IV*.

## V. DISCUSSION

The results of this mapping study indicate that issues related to dynamic metrics are receiving increasing attention from researchers, with the number of publications increasing greatly from 2002 onwards. More than half the entire body of papers was published between 2007 and 2011.

In this section we address the review questions noted in *Section III*. We also discuss the limitations of our study.

### RQ1  Which aspects of dynamic metrics have been most frequently subjected to study?

The most widely studied aspects of dynamic metrics have been software complexity, memory allocation and usage, and code execution metrics. A relatively large proportion of research has been focused on coupling, cohesion and other complexity and maintainability metrics. Coupling has been the most studied single metric type. A number of studies have proposed new or amended sets of coupling-related metrics. Several empirical studies used dynamic coupling metrics to collect data for the purpose of software comprehension and/or re-engineering.

### RQ2  Which aspects of dynamic metrics could be recommended as topics for future research?

In respect to the many dimensions of software quality, it is clear that complexity- and maintainability-oriented dynamic metrics have been the most widely discussed in the literature; however the same high level of attention has not been directed to metrics for other quality dimensions, such as testability and reusability. In our view dynamic metrics could be well suited to measure, and predict, testability. Measuring testability dynamically could be effective, particularly when considering different levels of testing (e.g., unit, integration, system) and the relationships between components. In addition, we considered the use of dynamic metrics at various levels of development and found that the proportion of investigations into design-level metrics is relatively low when compared to that for code-level studies. If useful metrics can be determined at the design stage this could help to reduce or minimize the risk of later costly failures.

## VI. VALIDITY THREATS

The main validity threat to this review study is the incomplete or inappropriate selection of publications. In spite of us following a systematic approach, it is still possible that we have missed some relevant studies especially if they were published in sources other than those we considered or that had not been cited in any of the articles selected in our search. To mitigate this risk we defined our search string for the automatic search alongside the search strategy. However, it is still possible that we have missed some relevant papers. There is also a chance that some related papers have used terms other than those we used in our search string. If terms other than "dynamic metrics", "dynamic measurement", or "runtime metrics" were used then the possibility of us missing a study is high. To avoid such a problem we repeatedly refined our search string and performed sequential testing in order to recognize and include as many relevant studies as possible. In addition, we conducted reference checks on all reference lists of the selected articles on the topic, to locate any missing influential articles. The selected studies were then examined and subsequently added to the final set of papers to be reviewed. In our view this was of use in limiting the number of missing influential articles (although we are unable to 'prove' this). Furthermore, the manual search we conducted was intended to fill any gaps by directly targeting relevant reputable publishing venues.

TABLE IV. SUMMARY OF CURRENT DYNAMIC METRICS TOOLS

| Tool | Description |
|---|---|
| *J [9] | *J is a tool designed to ease the dynamic metrics data collection process. The tool is used to gather, compute and present dynamic metrics data for Java programs. |
| A new debugging tool [72] | A debugging tool to compute the number of executions for individual methods. This is the only tool found that has been designed specifically for the C language. |
| AOP Hidden-Metrics [27] | An AOP-based adaptable tool that collects dynamic metric data in a non-invasive way. They employ an AOP technique using the AspectJ tool. |
| *Senseo* [17] | A plug-in to enhance the traditional static information provided by Eclipse with various dynamic metrics information. *Senseo* collects both runtime information and performance-related metrics. |
| DynaRIA [66] | A tool designed to support and enhance the comprehension of *Ajax* applications for the purpose of maintenance and testing. |
| A new (CCRCs) profiler [59] | A profiler that uses (CCRCs) visualization charts to enable efficient construction and navigation of large *Calling Context Trees* during execution. It also provides a visualization environment for the collected dynamic data. |

## VII. CONCLUSIONS

We have conducted and reported a systematic mapping study of dynamic software metrics. An automatic search was used to look for articles using the *SCOPUS* and *Google Scholar* search engines. We also conducted a manual search to look for articles that could have been missed by the automatic search. In addition we carried out reference checking to maximize article coverage and minimize the chances of omitting significant articles.

Out of all the published articles we scanned we identified 60 papers related to dynamic metrics. Following that we categorized the articles based on their Research focus, type and contribution. The results of this study indicate that there is growing interest in dynamic metrics in the software engineering research community. Publications have noticeably increased in number, particularly from the year 2002 onward. More than half of the articles selected were published from 2007 onward.

We conclude that most research to date has addressed complexity- and maintainability-related aspects of dynamic metrics. A great deal of measurement focus has been given to factors such as coupling, cohesion and polymorphism. This may be due to the fact that such aspects are the 'low-hanging fruit' – but whether they have actual value in industry is not yet established. Very little attention has been directed to other quality-related factors such as testability and reusability. Moreover, we also note that metrics for OO systems dominate the literature. Metrics for Procedural, Web, Aspect-Oriented and other systems are far fewer in number.

Our future research plan as informed by this study includes wider investigation and analysis of dynamic software metrics. In particular, we plan to investigate the feasibility of using dynamic metrics to measure other software quality characteristics and to empirically assess the cost-benefit of dynamic metric collection and analysis.

## REFERENCES

[1] J. K. Chhabra and V. Gupta, "A survey of dynamic software metrics," *J. Comput. Sci. Technol.,* vol. 25, pp. 1016-1029, 2010.

[2] O. Gómez, H. Oktaba, M. Piattini, and F. García, "A Systematic Review Measurement in Software Engineering: State-of-the-Art in Measures," in *Communications in Computer and Information Science- Software and Data Technologies*. vol. 10, J. Filipe*, et al.*, Eds., ed: Springer Berlin Heidelberg, 2008, pp. 165-176.

[3] B. Kitchenham, "What's up with software metrics? - A preliminary mapping study," *Journal of Systems and Software,* vol. 83, pp. 37-51, 2010.

[4] B. Cornelissen, A. Zaidman, A. v. Deursen, L. Moonen, and R. Koschke, "A Systematic Survey of Program Comprehension through Dynamic Analysis," *IEEE Transactions on Software Engineering,* vol. 35, pp. 684-702, 2009.

[5] T. Ball, "The concept of dynamic analysis," *SIGSOFT Software Engineering Notes,* vol. 24, pp. 216-234, 1999.

[6] M. D. Ernst, "Static and dynamic analysis: Synergy and duality," in *ACM SIGPLAN-SIGSOFT Workshop on Program Analysis for Software Tools and Engineering*, Portland, USA, 2003, pp. 24 - 27.

[7] E. Arisholm, L. C. Briand, and A. Foyen, "Dynamic Coupling Measurement for Object-Oriented Software," *IEEE Transactions on Software Engineering,* vol. 30, pp. 491-506, 2004.

[8] J. Cleland-Huang, C. K. Chang, H. Kim, and A. Balakrishnan, "Requirements-based dynamic metrics in object-oriented systems," in *International Symposium on Requirements Engineering*, 2001, p. 212.

[9] B. Dufour, K. Driesen, L. Hendren, and C. Verbrugge, "Dynamic metrics for java," in *ACM SIGPLAN Conference on Object-Oriented Programing, Systems, Languages, and Applications*, Anaheim, California, USA, 2003, pp. 149-168.

[10] A. Mitchell and J. F. Power, "Using object-level run-time metrics to study coupling between objects," in *ACM Symposium on Applied Computing*, Santa Fe, New Mexico, 2005, pp. 1456-1462.

[11] S. M. Yacoub, H. H. Ammar, and T. Robinson, "Dynamic metrics for object oriented designs," in *International Symposium on Software Metrics*, 1999, p. 50.

[12] H. Pirzadeh, A. Agarwal, and A. Hamou-Lhadj, "An Approach for Detecting Execution Phases of a System for the Purpose of Program Comprehension," in *International Conference on Software Engineering Research, Management and Applications*, 2010, pp. 207-214.

[13] B. Cornelissen, A. Zaidman, and A. v. Deursen, "A Controlled Experiment for Program Comprehension through Trace Visualization," *IEEE Transactions on Software Engineering,* vol. 37, pp. 341-355, 2011.

[14] E. Safari-Sharifabadi and C. Constantinides, "Dynamic analysis of Ada programs for comprehension and quality measurement," in

*SIGAda Annual International Conference*, Portland, OR, USA, 2008, pp. 15-38.

[15] T. Richner and S. Ducasse, "Recovering high-level views of object-oriented applications from static and dynamic information," in *IEEE International Conference on Software Maintenance*, 1999, p. 13.

[16] M. J. Pacione, M. Roper, and M. Wood, "A comparative evaluation of dynamic visualisation tools," in *Working Conference on Reverse Engineering*, 2003, pp. 80-89.

[17] D. Rothlisberger*, et al.*, "Augmenting static source views in IDEs with dynamic metrics," in *International Conference on Software Maintenance*, 2009, pp. 253-262.

[18] V. Gupta and J. K. Chhabra, "Dynamic cohesion measures for object-oriented software," *Journal of Systems Architecture,* vol. 57, pp. 452-462, 2011.

[19] B. Adams*, et al.*, "Using aspect orientation in legacy environments for reverse engineering using dynamic analysis-An industrial experience report," *Journal of Systems and Software,* vol. 82, pp. 668-684, 2009.

[20] E. Stroulia and T. Systä, "Dynamic analysis for reverse engineering and program understanding," *ACM SIGAPP Applied Computing Review,* vol. 10, pp. 8-17, 2002.

[21] E. S. Cho, C. J. Kim, S. D. Kim, and S. Y. Rhew, "Static and dynamic metrics for effective object clustering," in *Asia Pacific Software Engineering Conference*, 1998, p. 78.

[22] A. Zaidman and S. Demeyer, "Automatic identification of key classes in a software system using webmining techniques," *Journal of Software Maintenance and Evolution: Research and Practice* vol. 20, pp. 387-417, 2008.

[23] J. Voas, "Dynamic testing complexity metric," *Software Quality Journal,* vol. 1, pp. 101-114, 1992.

[24] J. C. Munson and T. M. Khoshgoftaar, "Measuring dynamic program complexity," *IEEE Software,* vol. 9, pp. 48-55, 1992.

[25] B. Dufour, L. Hendren, and C. Verbrugge, "*J: a tool for dynamic analysis of Java programs," in *ACM SIGPLAN Conference on Object-Oriented Programing, Systems, Languages, and Applications*, Anaheim, CA, USA, 2003, pp. 306-307.

[26] B. Dufour*, et al.*, "Measuring the dynamic behaviour of AspectJ programs," in *Annual ACM SIGPLAN Conference on Object-Oriented Programming, Systems, Languages, and Applications*, Vancouver, BC, Canada, 2004, pp. 150-169.

[27] W. Cazzola and A. Marchetto, "AOP-HiddenMetrics: Separation, Extensibility and Adaptability in SW Measurement," *Journal of Object Technology,* vol. 7, pp. 53–68, 2008.

[28] B. Kitchenham and S. Charters, "Guidelines for performing systematic literature reviews in software engineering," Keele University and University of Durham, Technical Report 2007.

[29] B. A. Kitchenham, D. Budgen, and O. Pearl Brereton, "Using mapping studies as the basis for further research – A participant-observer case study," *Information and Software Technology,* vol. 53, pp. 638-651, 2011.

[30] S. MacDonell, M. Shepperd, B. Kitchenham, and E. Mendes, "How reliable are systematic reviews in empirical software engineering?," *IEEE Transactions on Software Engineering,* vol. 36, pp. 676-687, 2010.

[31] B. Kitchenham*, et al.*, "Refining the systematic literature review process—two participant-observer case studies," *Empirical Software Engineering,* vol. 15, pp. 618-653, 2010.

[32] K. Petersen, R. Feldt, S. Mujtaba, and M. Mattsson, "Systematic Mapping Studies in Software Engineering," in *International Conference on Evaluation and Assessment in Software Engineering*, University of Bari, Italy, 2008.

[33] R. Wieringa, N. Maiden, N. Mead, and C. Rolland, "Requirements engineering paper classification and evaluation criteria: a proposal and a discussion," *Requirements Engineering,* vol. 11, pp. 102-107, 2006.

[34] S. M. Yacoub and H. H. Ammar, "A Methodology for Architecture-Level Reliability Risk Analysis," *IEEE Transaction on Software Engineering,* vol. 28, pp. 529-547, 2002.

[35] A. Mitchell and J. F. Power, "A study of the influence of coverage on the relationship between static and dynamic coupling metrics," in *Science of Computer Programming*, 2006, pp. 4-25.

[36] A. Mitchell and J. F. Power, "Run-time Coupling Metrics for the Analysis of Java Programs - preliminary results from the SPEC and Grande suites," Department of Computer Science, National University of Ireland, Technical Report 2003.

[37] Á. Mitchell and J. F. Power, "An empirical investigation into the dimensions of run-time coupling in Java programs," presented at the Proceedings of the 3rd international symposium on Principles and practice of programming in Java, Las Vegas, Nevada, 2004.

[38] A. Mitchell and J. F. Power, "Toward a definition of run-time object-oriented metrics," presented at the Workshop on Quantitative Approaches in Object-Oriented Software Engineering, Darmstadt, Germany, 2003.

[39] A. Zaidman and S. Demeyer, "Analyzing large event traces with the help of coupling metrics," presented at the International Workshop on Object-Oriented Reengineering, Antwerp, Belgium, 2004.

[40] R. Gunnalan, M. Shereshevsky, and H. H. Ammar, "Pseudo dynamic metrics [software metrics]," presented at the International Conference on Computer Systems and Applications, 2005.

[41] Y. Hassoun, S. Counsell, and R. Johnson, "Dynamic coupling metric: proof of concept," *IEE Proceedings -Software,* vol. 152, pp. 273-279, 2005.

[42] A. Beszedes, T. Gergely, S. Farago, T. Gyimothy, and F. Fischer, "The Dynamic Function Coupling Metric and Its Use in Software Evolution," presented at the European Conference on Software Maintenance and Reengineering, 2007.

[43] P. T. Quynh and H. Q. Thang, "Dynamic Coupling Metrics for Service-Oriented Software," *International Journal of Computer Science and Engineering,* vol. 3, p. 6, 2009.

[44] S. Allier, S. Vaucher, B. Dufour, and H. Sahraoui, "Deriving Coupling Metrics from Call Graphs," in *International Working Conference on Source Code Analysis and Manipulation*, 2010, pp. 43-52.

[45] S. G. Maisikeli and F. J. Mitropoulos, "Aspect mining using Self-Organizing Maps with method level dynamic software metrics as input vectors," in *International Conference on Software Technology and Engineering*, 2010, pp. V1-212-V1-217.

[46] A. Tahir, R. Ahmad, and Z. M. Kasirun, "Maintainability dynamic metrics data collection based on aspect-oriented technology," *Malaysian Journal of Computer Science,* vol. 23, pp. 177-194, 2010.

[47] S. Babu and R. M. S. Parvathi, "Design dynamic coupling measurement of distributed object oriented software using trace events," *Journal of Computer Science,* vol. 7, pp. 770-778, 2011.

[48] R. Burrows, F. Taïani, A. Garcia, and F. C. Ferrari, "Reasoning about Faults in Aspect-Oriented Programs: A Metrics-Based Evaluation," in *International Conference on Program Comprehension*, 2011, pp. 131-140.

[49] R. Burrows, F. C. Ferrari, A. Garcia, and F. Taïani, "An empirical evaluation of coupling metrics on aspect-oriented programs," in *Proceedings of the 2010 ICSE Workshop on Emerging Trends in Software Metrics*, Cape Town, South Africa, 2010, pp. 53-58.

[50] N. Gupta and P. Rao, "Program Execution-Based Module Cohesion Measurement," in *International Conference on Automated Software Engineering*, 2001, p. 144.

[51] A. Mitchell and J. F. Power, "Run-Time Cohesion Metrics: An Empirical Investigation," in *International Conference on Software Engineering Research and Practice*, 2004.

[52] P. Khurana and P. J. Kaur, "Dynamic Metrics at design Level," *International Journal of Information Technology and Knowledge Management,* vol. 2, pp. 449-454, 2009.

[53] J. C. Munson and G. A. Hall, "Estimating test effectiveness with dynamic complexity measurement," *Empirical Software Engineering,* vol. 1, pp. 279-305, 1996.

[54] K. J. Keen, R. Mathur, and L. Etzkorn, "Towards a measure of software intelligence employing a runtime complexity metric," presented at the International Conference on Software Engineering and Applications, Cambridge, MA, USA, 2009.

[55] R. Mathur, K. J. Keen, and L. H. Etzkorn, "Towards an object-oriented complexity metric at the runtime boundary based on decision points in code," presented at the Annual Southeast Regional Conference, Oxford, Mississippi, 2010.

[56] W. Yuying, L. Qingshan, C. Ping, and R. Chunde, "Dynamic Fan-in and Fan-out Metrics for Program Comprehension," presented at the International Workshop on Program Comprehension through Dynamic Analysis, Pittsburgh, Pennsylvania, USA, 2005.

[57] W. Binder, J. Hulaas, and P. Moret, "Reengineering Standard Java Runtime Systems through Dynamic Bytecode Instrumentation," presented at the International Working Conference on Source Code Analysis and Manipulation, 2007.

[58] W. Binder, *et al.*, "Towards a domain-specific aspect language for dynamic program analysis," in *Annual Workshop on Domain-Specific Aspect Languages*, 2011, pp. 9-11.

[59] P. Moret, W. Binder, A. Heydarnoori, and D. Ansaloni, "Tool demonstration: effective runtime exploration of the inter-procedural control flow in Java applications," in *International Conference on the Principles and Practice of Programming in Java*, Vienna, Austria, 2010, pp. 162-165.

[60] D. Rothlisberger, "Exploiting Dynamic Information in IDEs Eases Software Maintenance," in *International Workshop on Program Comprehension through Dynamic Analysis*, Massachusetts, USA, 2010.

[61] H. Gani, C. Ryan, and P. Rossi, "Runtime metrics collection for middleware supported adaptation of mobile applications," presented at the Workshop on Adaptive and Reflective Middleware, Melbourne, Australia, 2006.

[62] B. Dufour, L. Hendren, and C. Verbrugge, "Problems in Objectively Quantifying Benchmarks using Dynamic Metrics," Sable Research Group, School of Computer Science, McGill University, Technical Report 2003.

[63] K. H. T. Choi and E. Tempero, "Dynamic measurement of polymorphism," in *Australasian Conference on Computer Science*, Ballarat, Victoria, Australia, 2007, pp. 211-220.

[64] P. S. Sandhu and G. Singh, "Dynamic Metrics for Polymorphism in Object Oriented Systems," *World Academy of Science, Engineering and Technology,* vol. 39, 2008.

[65] A. Mitchell and J. F. Power, "An approach to quantifying the run-time behaviour of Java GUI applications," presented at the Winter International Synposium on Information and Communication Technologies, Cancun, Mexico, 2004.

[66] D. Amalfitano, A. R. Fasolino, A. Polcaro, and P. Tramontana, "DynaRIA: A Tool for Ajax Web Application Comprehension," presented at the International Conference on Program Comprehension, 2010.

[67] E. Mendes, N. Mosley, and S. Counsell, "Investigating Web size metrics for early Web cost estimation," *Journal Systems and Software,* vol. 77, pp. 157-172, 2005.

[68] S. R. Chidamber and C. F. Kemerer, "A metrics suite for object oriented design," *IEEE Transactions on Software Engineering* vol. 20, pp. 476-493, 1994.

[69] R. Mathur, K. J. Keen, and L. H. Etzkorn, "Towards an object-oriented complexity metric at the runtime boundary based on decision points in code," in *Annual Southeast Regional Conference*, Oxford, Mississippi, 2010, pp. 1-5.

[70] Y. Cai, "Assessing the Effectiveness of Software Modularization Techniques through the Dynamics of Software Evolution," in *Workshop on Assessment of Contemporary Modularization Techniques*, Orlando, US, 2008.

[71] V. L. Narasimhan and B. Hendradjaya, "Some theoretical considerations for a suite of metrics for the integration of software components," *Information Sciences,* vol. 177, pp. 844-864, 2007.

[72] K. Aggarwal, Y. Singh, and J. Chhabra, "A dynamic software metric and debugging tool," *SIGSOFT Software Engineering Notes,* vol. 28, p. 1, 2003.